\documentclass[conference,twocolumn]{IEEEtran}

\ifCLASSINFOpdf
\else
\fi
\usepackage{hyperref}
\usepackage{setspace}
\usepackage{graphicx}
\usepackage{amsthm}
\usepackage{subfigure}

\newtheorem{Lemma}{Lemma}

\begin{document}

\title{Opportunistic Adaptive Relaying in Cognitive Radio Networks}

\author
{\IEEEauthorblockN{Wael Jaafar}
\IEEEauthorblockA{\'{E}cole Polytechnique de Montr\'{e}al\\Department of Electrical Engineering\\
Montreal, Canada\\
Email: wael.jaafar@polymtl.ca}
\and
\IEEEauthorblockN{Wessam Ajib}
\IEEEauthorblockA{Universit\'{e} du Qu\'{e}bec \`{a} Montr\'{e}al\\
Department of Computer Science\\Montreal, Canada\\
Email: ajib.wessam@uqam.ca}
\and
\IEEEauthorblockN{David Haccoun}
\IEEEauthorblockA{\'{E}cole Polytechnique de Montr\'{e}al\\
Department of Electrical Engineering\\
Montreal, Canada\\
Email: david.haccoun@polymtl.ca} }

\maketitle

\begin{abstract}
Combining cognitive radio technology with user cooperation could be advantageous to both primary and secondary  transmissions. In this paper, we propose a first relaying scheme for cognitive radio networks (called ``Adaptive relaying scheme~1"), where one relay node can assist the primary or the secondary transmission with the objective of improving the outage probability of the secondary transmission with respect to a
primary outage probability threshold. Upper bound expressions of the secondary outage probability using the proposed scheme are derived over Rayleigh fading channels. Numerical and simulation results show that the secondary outage probability using the proposed scheme is lower than that of other relaying schemes. Then, we extend the proposed scheme to the case where the relay node has the ability to decode both the primary and secondary signals and also can assist simultaneously both transmissions. Simulations show the performance improvement that can be obtained due to this extension in terms of secondary outage probability.
\end{abstract}

\IEEEpeerreviewmaketitle

\section{Introduction}
In order to overcome the problems related to the rigid allocation of spectrum bands to few licensed operators
and the under-utilization of these bands, Cognitive
Radio (CR) technology has evolved in wireless communications for allowing unlicensed secondary users (SUs) to access licensed parts of the spectrum without harmfully interfering with the transmissions of the licensed primary users (PUs) taking place in the same spectrum band \cite{Mitola}-\cite{Haykin}. In underlay spectrum sharing mode, SUs are allowed to transmit simultaneously with PUs if they tune their transmission parameters (such as
transmit power) to be harmless to primary transmissions. 

Meanwhile, user cooperation has been recognized as an interesting technique that allows to achieve increased diversity order when one or several relay nodes assist the transmission \cite{Laneman1}-\cite{Jaafar_WCM}.

Consequently, combining user cooperation and cognitive radio has
recently attracted attention to improve both the spectrum utilization and
the transmission performances. User cooperation has been applied for the primary transmission when a
cognitive secondary transmitter acts as a relay \cite{Yang_2009}.
By doing so, the primary outage probability is improved, while SUs have more
opportunities to access the licensed spectrum bands and hence secondary transmission performances can be also improved. In
\cite{Zou}, an adaptive user cooperation scheme with best-relay selection is proposed in multiple-relay Cognitive Radio Networks (CRNs) to improve the secondary outage performance while satisfying a primary outage probability threshold. By letting the ``best" CR relay assist the secondary transmission, the secondary outage probability can be considerably reduced.
In \cite{Yener}, secondary transmissions are assisted by a group of CR relay nodes located at different positions. The outage probability was investigated when all the relays forward simultaneously their received signals. The system achieves full diversity when the number of cooperating relay nodes is adequately selected. In \cite{Jaafar_Globecom}, we proposed a new relaying scheme for CRNs where a CR relay node is able to assist
simultaneously the primary and secondary transmissions. It has been shown that for certain relay's position, assisting simultaneously both transmissions provides better outage performance than assisting only the primary or the secondary transmission. However, the proposed scheme is greedy on the relay's transmit power.

Even though, the previous works have investigated the utilization of user cooperation in CRNs for assisting the
primary transmissions, the secondary transmissions or both, no work has investigated adaptive relaying schemes where the relay node decides independently when and which communication to assist. Consequently, we propose in this paper a novel opportunistic adaptive relaying scheme, where the CR relay node is able to decide when to cooperate or not, and in case of cooperation, whom to cooperate with (primary transmission or secondary transmission) depending on the channel states. We propose to extend the adaptive relaying scheme to the case where the relay node can also cooperate simultaneously with both transmissions.

The paper is organized as follows. Section II presents the system
model. In section III, we describe the two adaptive relaying schemes and we provide analytically the secondary outage probability for the first adaptive relaying scheme. Section IV shows and discusses the numerical and simulation results and a conclusion closes the paper in section V.

\section{System Model}

\begin{figure}[t]
  \centering
  \includegraphics[width=220pt]{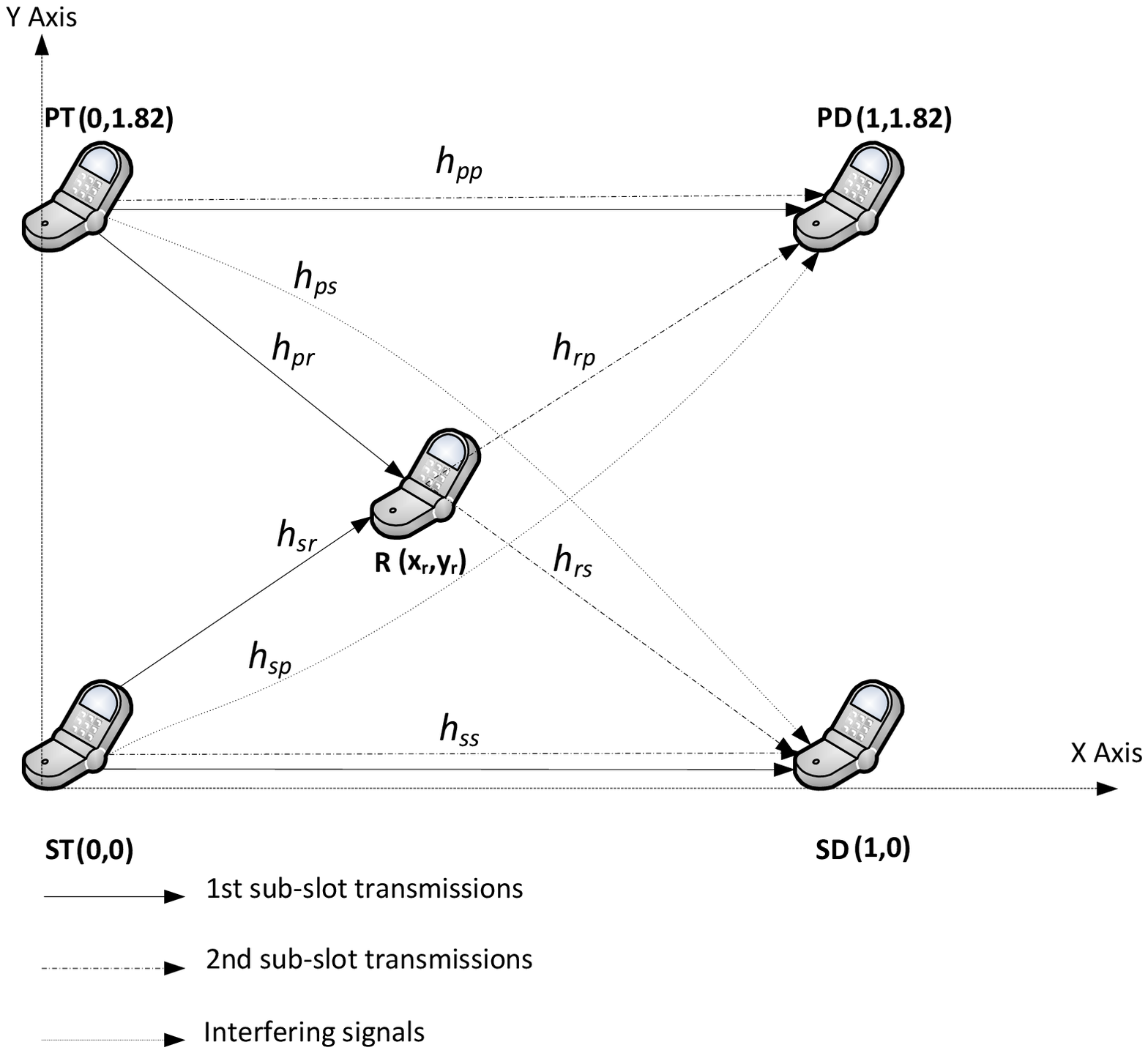}
  \caption{The Cognitive Radio Network}
  \label{Fig:Network}
\end{figure}

We assume a CRN where one primary transmitter (PT)
transmits data to a primary destination (PD) and a secondary
transmitter (ST) communicates with a secondary destination (SD)
over the same frequency band (Fig. \ref{Fig:Network}). We assume a decode-and-forward secondary CR relay node (R) that can assist the primary or the secondary transmission in order to increase the secondary access to the licensed spectrum bands with respect to a certain primary outage probability threshold. In the extension, we assume that the relay is able to assist both transmissions.

We assume that PT and ST transmit their signals $x_p$ and $x_s$
(where $E\{|x_p|^2\}=E\{|x_s|^2\}=1$) with powers $P_{p}$ and
$P_{s}$ respectively in order to achieve data rates $R_p$ and
$R_s$ respectively. We assume also that R uses transmit power $P_r
\leq P_r^{max}$, where $P_r^{max}$ is the maximal relay transmit power.
We assume that the channels are submitted to Rayleigh fading
and path loss attenuation and are stationary during a time-slot (time slot $=1^{st}$
+ $2^{nd}$ sub-slots).

Following Fig. \ref{Fig:Network}, the received signals during the first sub-slot are expressed
by:
\begin{equation}
\label{eq:received_PD_direct}
y_{a}(1)=\sqrt{P_{p}}h_{pa}x_p+\sqrt{P_{s}}h_{sa}x_s+n_{a},
\end{equation}
where $a=p,s$ or $r$ ($p$, $s$ and $r$ denote primary, secondary
and relay node resp.), $h_{ba}$ ($b=p$ or $s$) is the channel gain between nodes $b$ and $a$ having variance $\sigma_{ba}^{2}=d_{ba}^{-\beta}$ , $d_{ba}$ is the distance between $b$ and $a$, $\beta$ is the path-loss exponent, and where $n_{a}$ is the
additive white gaussian noise with zero mean and variance
$N_0$ received at $a$. We assume a fixed $P_p$ and that $P_s$ is calculated with respect to the
primary outage probability threshold denoted $\varepsilon$.  $P_s$ is
given similarly to \cite{Zou} by:
\begin{equation}
\label{eq:condition_gamma_ST}
P_s=\frac{2 P_p \sigma_{pp}^{2}}{\Lambda_p \sigma_{sp}^{2}} \rho^{+},
\end{equation}
where $\rho^{+}=max(0,\rho)$, $\rho=\frac{e^{-\frac{\Lambda_p}{2
{\gamma}_{p}\sigma_{pp}^{2}}}}{1-\varepsilon}-1$, $\Lambda_p=2^{2 R_p}-1$ and $\gamma_p = P_p/N_0$. ST calculates $P_s$ assuming that PT repeats the same signal over the two sub-slots with the same transmit power $P_p$.

\section{Adaptive Relaying Schemes}

\subsection{Description of adaptive relaying scheme 1}

This novel scheme aims to exploit efficiently the relay position, the acquired information and the
propagation environment conditions. We define by $a_0=\frac{\gamma_s |h_{ss}|^2}{\gamma_p |h_{ps}|^2+1}$, $a_p=\frac{\gamma_s |h_{ss}|^2}{\gamma_r^{(p)} |h_{rs}|^2+1}$, $a_s=\frac{\gamma_r^{(s)} |h_{rs}|^2}{\gamma_p |h_{ps}|^2+1}$ where $\gamma_a=P_a/N_0$ ($a=p$ or $s$) and $\gamma_{r}^{(i)}$ is the transmit power of the relay node when assisting transmission $i$. We also define $E_i=\left\{ a_i=max\left\{ a_p, a_s, a_0 \right\} \right\}$ the opportunism condition ($i=p, s$ or $0$).
At the first sub-slot, R attempts to decode $x_p$ or $x_s$  and then, a relaying procedure is chosen depending on the value of $D$ defined by:
\begin{eqnarray}
\label{eq:condition_CR_helps_PU_SU}
\nonumber &\mathrm{If }\;A_p \cap \left\{ \left\{\bar{A_s} \cap (a_p > a_0)\right\} \cup \left\{A_s \cap E_p \right\} \right\} \mathrm{, then} &D = 1,\\
\nonumber &\mathrm{If }\;A_s \cap \left\{ \left\{\bar{A_p} \cap (a_s > a_0)\right\} \cup \left\{A_p \cap E_s \right\} \right\} \mathrm{, then} &D = 2,\\
\nonumber &\mathrm{Otherwise} &D = 0,\\
\end{eqnarray}
where $\bar{A}$ is the complement of $A$, and
\begin{equation}
A_p = \left\{\frac{1}{2}{{\log }_2}\left( {1 + \frac{{\gamma _p}{{\left| {h_{pr}} \right|}^2}}{{\gamma _s}{{\left| {h_{sr}} \right|}^2}+1}} \right){\rm{ \geq }}{R_p}\right\},
\end{equation}
\begin{equation}
A_s = \left\{\frac{1}{2}{{\log }_2}\left( {1 + \frac{{\gamma _s}{{\left| {h_{sr}} \right|}^2}}{{\gamma _p}{{\left| {h_{pr}} \right|}^2}+1}} \right){\rm{ \geq }}{R_s}\right\}.
\end{equation}
The comparison of $(a_i: i = p, s$ or $0)$ indicates which relaying would improve better the secondary outage probability.

The different cases are detailed below.

\subsubsection{R assists the primary transmission ($D=1$)}

This case occurs either when (i) R succeeds to decode the primary signal but not the secondary signal and when relaying the primary signal provides lower secondary outage probability than the repetition (i.e., $a_p > a_0$) or (ii) R succeeds to decode both the primary and secondary signals and assisting the primary transmission provides the lowest $P_{out_{sec}}$(i.e., $E_p$). Hence, when the relay is able to decode the primary signal and the best choice is to assist the primary transmission, then $D=1$. Consequently, the received signals at PD and SD, on the second sub-slot, are respectively given by:
\begin{equation}
\label{eq:received_R_helpPU_D1}
y_{a}(2|D=1)=\sqrt{P_{r}^{(p)}}h_{ra}x_p+\sqrt{P_{s}}h_{sa}x_s+n_{a}.
\end{equation}
After normalizing the noise variances and combining the signals received in the two sub-slots (given by
(\ref{eq:received_PD_direct}) and (\ref{eq:received_R_helpPU_D1}))
with Maximum Ratio Combining (MRC) \cite{Yang_2009}, the
Signal-to-Interference-plus-Noise-Ratio (SINR) at PD is:
\begin{equation}
\label{eq:SINR_PD_D1}
SINR_{p}(D=1)=\frac{\gamma_{p}|h_{pp}|^2}{\gamma_{s}|h_{sp}|^2+1}+\frac{\gamma_{r}^{(p)}|h_{rp}|^2}{\gamma_{s}|h_{sp}|^2+1},
\end{equation}
while SINR$_{s}$ at SD is given by:
\begin{equation}
\label{eq:SINR_SD_D1_CR_help_PU}
SINR_{s}(D=1)=\frac{\gamma_{s}|h_{ss}|^2}{\gamma_{p}|h_{ps}|^2+1}+\frac{\gamma_{s}|h_{ss}|^2}{\gamma_{r}^{(p)}|h_{rs}|^2+1}.
\end{equation}

\subsubsection{R assists the secondary transmission ($D=2$)}

This second case occurs when (i) R succeeds to decode only the secondary signal and when relaying the secondary signal is beneficial to the secondary transmission (i.e., $a_s > a_0$) or (ii) R succeeds to decode both signals and assisting the secondary transmission provides the lowest $P_{out_{sec}}$(i.e., $E_s$). In this case, R assists the secondary transmission and $D=2$. The received SINR at PD and that at SD are expressed as (\ref{eq:SINR_SD_D1_CR_help_PU}) and (\ref{eq:SINR_PD_D1}) respectively, where indexes $p$ and $s$ are inverted.
\subsubsection{R does not assist the transmissions ($D=0$)}

When the relay is not able to decode the signals or when relaying is not beneficial to $P_{out_{sec}}$, the relay does not participate in the transmissions. In this case, we assume that the primary and secondary transmitters retransmit the same signals. Accordingly, the received SINR at SD is given by (eq.(7), \cite{Jaafar_Globecom}) and that at PD by inverting indexes $p$ and $s$ in (eq.(7), \cite{Jaafar_Globecom}).

\subsection{Outage Probability Analysis}
In this scheme, any of three cases can happen. We start by calculating the probability of occurrence of each one.

An upper bound on the probability of occurrence of $D=1$ is given by:
\begin{eqnarray}
\label{eq:occurrence_D4_1}
\nonumber P\left(D=1\right)&=& \frac{\tilde{\gamma}_{ps}}{\tilde{\gamma}_{ps}+\tilde{\gamma}_{rs}^{(p)}} \times \frac{\tilde{\gamma}_{pr}e^{-\frac{\Lambda_p}{\tilde{\gamma}_{pr}}-\frac{\Lambda_s(1+\Lambda_p)}{(1-\Lambda_p\Lambda_s)\tilde{\gamma}_{sr}}}}{\tilde{\gamma}_{pr}+\Lambda_p \tilde{\gamma}_{sr}}\\
\nonumber &+& \frac{\tilde{\gamma}_{ps}}{\tilde{\gamma}_{ps}+\tilde{\gamma}_{rs}^{(p)}} \left(1-\frac{\tilde{\gamma}_{sr}e^{-\frac{\Lambda_s}{\tilde{\gamma}_{sr}}}}{\tilde{\gamma}_{sr}+\Lambda_s \tilde{\gamma}_{pr}} \right)\\
\nonumber &\times & \left( 1-e^{\frac{-\Lambda_s(1+\Lambda_p)}{(1-\Lambda_p \Lambda_s)\tilde{\gamma}_{sr}}} \right)-\phi_i\left( -\frac{1}{\tilde{\gamma}_{ps}}- \frac{\tilde{\gamma}_{rs}^{(s)}}{\tilde{\gamma}_{ss}\tilde{\gamma}_{ps}} \right)\\
\nonumber &\times & \frac{\tilde{\gamma}_{ps}}{\tilde{\gamma}_{ps}+\tilde{\gamma}_{rs}^{(p)}}  \left( \frac{1+\tilde{\gamma}_{ps}}{\tilde{\gamma}_{ss}} \right)e^{ \frac{1}{\tilde{\gamma}_{ps}} + \frac{\tilde{\gamma}_{rs}^{(s)}}{\tilde{\gamma}_{ss}\tilde{\gamma}_{ps}} }\\
&\times & \frac{\tilde{\gamma}_{sr}e^{-\frac{\Lambda_s}{\tilde{\gamma}_{sr}}-\frac{\Lambda_p(1+\Lambda_s)}{(1-\Lambda_p\Lambda_s)\tilde{\gamma}_{pr}}}}{\tilde{\gamma}_{sr}+\Lambda_s \tilde{\gamma}_{pr}},
\end{eqnarray}
where $\tilde{\gamma}_{ab}=\gamma_a \sigma_{ab}^2$, $\tilde{\gamma}_{rb}^{(i)}=\gamma_r^{(i)} \sigma_{rb}^2$ ($a=p,s$ or $r$ and $b=p,s$ or $r$) and  $\phi_i(x)=\int_{-\infty}^{x}{\frac{e^t}{t}dt}$.

\begin{IEEEproof}
See Appendix A.
\end{IEEEproof}
By following similar calculations, we can obtain $P(D=2)$.
Finally, $P(D=0)=1-\sum\limits_{i=1}^{2}P(D=i)$.

We next present the conditional primary and conditional secondary outage probabilities for each case:
\begin{itemize}
  \item $D=1$
\end{itemize}
The conditional primary outage probability is given by:
\begin{eqnarray}
\label{eq:outage_prob_CR_help_PU_D1_0}
\nonumber P_{pri}(out.|D=1)&=&P\left( SINR_{p}(D=1)<\Lambda_p \right)\\
 &=&P(\omega<\Lambda_p+\Lambda_p \omega_1 - \omega_2)
\end{eqnarray}
where $\Lambda_a=2^{2R_a}-1$ ($a=p$ or $s$),
$\omega=\gamma_{r}^{(p)}|h_{rp}|^2$, $\omega_1=\gamma_{s}|h_{sp}|^2$ and
$\omega_2=\gamma_{p}|h_{pp}|^2$. We shall make use of the
following Lemma.
\begin{Lemma}
The exact closed-form expression of the conditional primary outage probability is expressed by:
\label{Lemma1}
\begin{equation}
\label{eq:outage_prob_CR_help_PU_D1} P_{pri}(out.|D=1)=\lambda_1+\lambda_2,
\end{equation}
\end{Lemma}
where
\begin{eqnarray}
\label{eq:d1}
\nonumber \lambda_1&=&\frac{\tilde{{\gamma}}_{sp}^2\Lambda_p^2
+\tilde{{\gamma}}_{rp}^{(p)}\tilde{{\gamma}}_{sp}\Lambda_p\left( 1-e^{-\frac{\Lambda_p}{\tilde{\gamma}_{rp}^{(p)}}} \right) }{(\tilde{{\gamma}}_{pp}+\Lambda_p
\tilde{{\gamma}}_{sp})(\tilde{{\gamma}}_{rp}^{(p)}+\Lambda_p \tilde{{\gamma}}_{sp})},\\
\\
\nonumber \lambda_2&=&\frac{\tilde{\gamma}_{pp}^2 \left( 1-e^{-\frac{\Lambda_p}{\tilde{\gamma}_{pp}}} \right)-\tilde{\gamma}_{pp} \tilde{\gamma}_{rp}^{(p)} \left( 1-e^{-\frac{\Lambda_p}{\tilde{\gamma}_{rp}^{(p)}}} \right) }{(\tilde{{\gamma}}_{pp}+\Lambda_p
\tilde{{\gamma}}_{sp})(\tilde{{\gamma}}_{rp}^{(p)}-\tilde{{\gamma}}_{pp})},\\
\label{eq:d2} &\forall&\; \tilde{{\gamma}}_{pp} \neq
\tilde{{\gamma}}_{rp}^{(p)}.
\end{eqnarray}
\begin{IEEEproof} See Appendix B.
\end{IEEEproof}
\begin{Lemma}
An upper bound on the conditional secondary outage probability is given by:
\label{Lemma2}
\begin{eqnarray}
\label{eq:conditional_sec_D1}
\nonumber P_{sec}(out.|D=1)&=& \left( 1-e^{-\frac{\Lambda_s}{\tilde{\gamma}_{ss}}}\left( 1+\frac{ln\left( 1+ \frac{\Lambda_s \tilde{\gamma}_{ps}}{\tilde{\gamma}_{ss}} \right)}{\tilde{\gamma}_{ps}} \right)  \right)\\
&\times& \left( \tilde{\gamma}_{rs}^{(p)} +1 \right)=\varphi \times \left( \tilde{\gamma}_{rs}^{(p)} +1 \right).
\end{eqnarray}
\end{Lemma}
\begin{IEEEproof} See Appendix C.
\end{IEEEproof}

By assisting the primary transmission, this relaying procedure aims to reduce the interference caused to the secondary transmission with respect to the primary outage threshold $\varepsilon$. For that purpose, R should control its transmit power $P_r$ to be as low as possible. This value of $P_r$, denoted $P_{r,num}^{(p)}$, is evaluated numerically by solving $P_{pri}(out.|D=1)=\varepsilon$. If $P_{r,num}^{(p)}>P_r^{max}$, then relaying is not beneficial and $D=0$.

\begin{itemize}
  \item $D=2$
\end{itemize}
Due to the similarity of the outage probability analysis of this relaying procedure to the first one, it is not given in details. However, by inverting indexes $p$ and $s$ and indexes $pri$ and $sec$ in equations (\ref{eq:outage_prob_CR_help_PU_D1_0})-(\ref{eq:conditional_sec_D1}), we obtain an accurate outage probability analysis. The relay transmit power should also be calculated such that $P_{pri}(out.|D=2)=\varepsilon$. The relay transmit power is given by
$\gamma_{r,num}^{(s)}=\frac{\varepsilon/\varphi'-1}{\sigma_{rp}^2}$, where
$\varphi'=\left( 1-e^{-\frac{\Lambda_p}{\tilde{\gamma}_{pp}}}\left( 1+\frac{ln\left( 1+ \frac{\Lambda_p \tilde{\gamma}_{sp}}{\tilde{\gamma}_{pp}} \right)}{\tilde{\gamma}_{sp}} \right)  \right)$.

Then, $\gamma_r^{(s)}=min\left( \gamma_{r,num}^{(s)}, \gamma_r^{max} \right)$.



\begin{itemize}
  \item $D=0$
\end{itemize}
The outage probability analysis of this case is presented in \cite{Jaafar_Globecom}. The conditional secondary outage probability is given by (eq.(16), \cite{Jaafar_Globecom}), while $P_{pri}(out.|D=0)$ is obtained by simply inverting indexes $p$ and $s$ in (eq.(16), \cite{Jaafar_Globecom}).

Finally, upper bounds on the primary and secondary outage probabilities are given by:
\begin{equation}
\label{eq:outage_adaptive}
P_{out_{c}}= \sum_{i=0}^{2}{P(D=i)P_c(out.|D=i)},
\end{equation}
where $c=pri$ or $sec$. The obtained expressions are upper bounds since some of the conditional outage probabilities calculated are upper bounds.

\subsection{Extension to ``Adaptive relaying scheme 2"}
In this extension, we assume that R is equipped with a SIC (Successive Interference Cancelation) receiver \cite{Tse}, and hence it is able to decode both signals.
We define $a_{ps}=\frac{(1-\alpha)\gamma_r^{(ps)} |h_{rs}|^2}{\alpha\gamma_r^{(ps)} |h_{rs}|^2+1}$, where $0 \leq \alpha \leq 1$.
We also define $C_{i}=\left\{a_i=max\left\{ a_0, a_p, a_s, a_{ps} \right\} \right\}$ ($i=0,p,s$ or $ps$) and $E=\left\{A_p \cap B_p \right\} \cup \left\{ A_s \cap B_s \right\}$ the event of a successful successive decoding of both signals, where $B_i = \left\{\frac{1}{2} \log_2\left( {1 + {\gamma _{i}}{{\left| {{h_{ir}}} \right|}^2}} \right){\rm{ \geq }}{R_i}\right\}$. We call this extension ``Adaptive relaying scheme 2". For this scheme, we distinguish four relaying procedures:
\begin{eqnarray}
\label{eq:condition_CR_helps_PU_SU}
\nonumber &\mathrm{If}\; E  \cap C_{ps},\;\mathrm{then}&\;D = 3,\\
\nonumber &\mathrm{If}\;\left\{A_s \cap \bar{B}_s \cap (a_s >a_0) \right\} \cup \left\{E \cap C_s \right\},\;\mathrm{then}&\;D = 2,\\
\nonumber &\mathrm{If}\;\left\{A_p \cap \bar{B}_p \cap (a_p >a_0)  \right\} \cup \left\{E \cap C_p \right\},\;\mathrm{then}&\;D = 1,\\
\nonumber &\mathrm{Otherwise}\;&D = 0.
\end{eqnarray}

The relaying procedures for $D=0,1$ or $2$ are identical to the first scheme. When $D=3$, a fraction of the relay transmit power $\alpha P_r^{(ps)}$ is used to send $x_p$ and the rest, i.e., $(1-\alpha) P_r^{(ps)}$ is used to transmit $x_s$. SINRs at PD and at SD are then given by (eq.(10), \cite{Jaafar_Globecom}) and (eq.(11), \cite{Jaafar_Globecom}) respectively. The parameter $\alpha$ is calculated using (eq.(33), \cite{Jaafar_Globecom}) where $\gamma_r^{(ps)}=\gamma_r^{max}$.


\begin{figure}
  \centering
  \includegraphics[width=225pt]{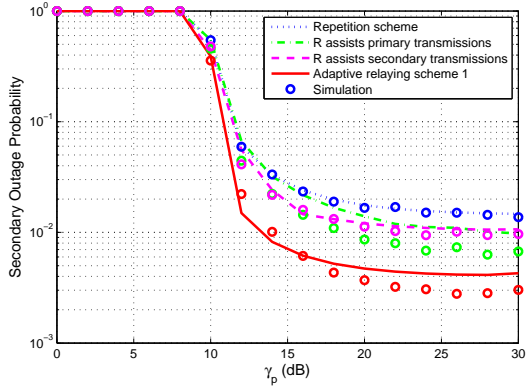}
  \caption{Comparaison of different relaying schemes}
  \label{Fig:Compare_schemes}
\end{figure}

\section{Numerical and Simulation Results}
We consider the CRN presented in Fig. \ref{Fig:Network} where the coordinates of PT, ST, PD and SD are given by (0,1.82), (0,0), (1,1.82) and (1,0) respectively (coordinates are in distance units). We assume that $R_p=0.8 bits/s/Hz$, $R_s=0.2 bits/s/Hz$, $\varepsilon=0.1$, $\beta=4$ and $\gamma_p=\gamma_r^{max}=20dB$. We measure the average of the secondary outage probability calculated for different random positions of the relay node in the plan of coordinates (X,Y) where $0.1 \leq X \leq 0.9$ and $0.1 \leq Y \leq 1.7$ unless otherwise is stated.
In Fig. \ref{Fig:Compare_schemes}, we compare the ``Adaptive
relaying scheme 1" to other transmission schemes presented in the literature \cite{Zou,Jaafar_Globecom}. At low $\gamma_p$ ($\gamma_p \leq 8dB$), no secondary transmissions are allowed. When $\gamma_p$ is higher than the cutoff value, the ``Adaptive relaying scheme 1" presents, as expected, the best outage performance since the proposed scheme chooses the most adequate relaying procedure. The ``R assists secondary transmissions" and ``R assists primary transmissions" schemes outperforms the non cooperative scheme. When R assists only the secondary transmissions, the performances are degraded by the fact that the relay transmit power is limited to $P_r^{max}$.


\begin{figure}
  \centering
  \includegraphics[width=225pt]{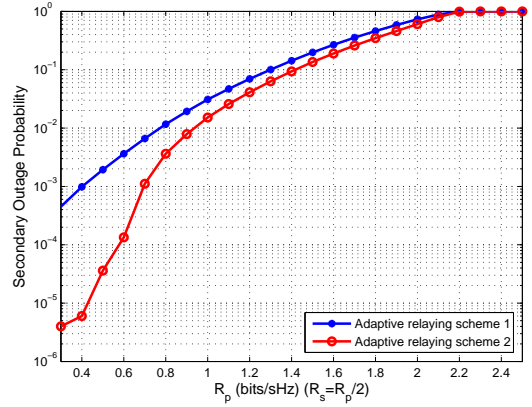}
  \caption{``Adaptive relaying scheme" versus ``Adaptive relaying scheme 2" (simulations only)}
  \label{Fig:compare_adapt12}
\end{figure}
\begin{figure}
  \centering
  \includegraphics[width=225pt]{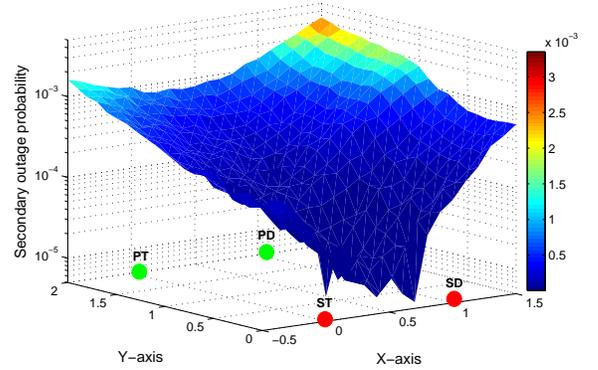}
  \caption{Secondary Outage Probability vs. Relay position (simulations only)}
  \label{Fig:3D}
\end{figure}

In Fig. \ref{Fig:compare_adapt12}, we compare the secondary outage probabilities of the two adaptive relaying schemes for different $R_p$ values and where we assume $R_s=\frac{R_p}{2}$. For $R_p \leq 2.2 bits/s/Hz$, the second scheme outperforms the first one. Indeed, the proposed second scheme offers more relaying possibilities for R and hence improves the secondary performance. For $R_p \geq 2.2 bits/s/Hz$, no secondary or relaying transmissions are allowed due to the high primary outage probability requirement that blocks any interfering transmission.

In Fig. \ref{Fig:3D}, we present the secondary outage performance using the
``Adaptive relaying scheme 2" for different positions of the relay
node on the plan (X,Y) where $-0.5 \leq X \leq 1.5$ and $0 \leq Y
\leq 2$. When R is close to the secondary nodes, the secondary outage performance is improved. Indeed, when R is close to the primary nodes, the cases $D=1$ and $D=3$ are predominant and the outage probability gain comes from the interference reduction. As R gets closer to the secondary nodes, the cases $D=2$ and $D=3$ occur more often. Hence, the outage performance gain is issued from the interference reduction ($D=3$) and cooperation ($D=2$). Moreover, when R is in the middle zone, the secondary outage probability decreases. In this zone, the condition of the channels linked to the relay node favors successful decoding and efficient forwarding of the signals.

\section{Conclusion}
In this paper, we proposed adaptive relaying schemes for cognitive radio networks, where a relay node is able to decide assisting the primary, the secondary or both transmissions depending on the channels condition. We showed by analysis and by simulation that the first adaptive relaying scheme, where the relay may help the primary or the secondary transmission, outperforms the non adaptive relaying schemes in terms of secondary outage probability, with respect to a primary outage probability threshold. Then, a second scheme has been proposed considering that the relay may help simultaneously both transmissions. Simulations show the performance improvement of the second scheme, specially at low data rates.

\appendices
\section{Proof of Eq.(\ref{eq:occurrence_D4_1})}
Due to the independency between events, we have:
\begin{eqnarray}
\label{eq:d_4_1_demo}
\nonumber P(D=1)&=&P\left( A_p \cap \bar{A}_s \right) P\left(a_p > a_0 \right)\\
\nonumber &+& P(A_p|A_s)P(A_s)P(E_p)\\
\nonumber &=&P(\gamma_p |h_{pr}|^2 \geq max\{ \Lambda_p (1+\gamma_s |h_{sr}|^2) ,\\
\nonumber &{}& \frac{\gamma_s |h_{sr}|^2}{\Lambda_s}-1\})P(\gamma_r^{(p)} |h_{rs}|^2 \leq \gamma_p |h_{ps}|^2)\\
\nonumber &+& P(\gamma_p |h_{pr}|^2\geq \frac{\Lambda_p (1+ \Lambda_s)}{1-\Lambda_p \Lambda_s})P(a_p>a_0)\\
&{}& P(\frac{\gamma_s |h_{sr}|^2}{\gamma_p |h_{pr}|^2+1}\geq \Lambda_s)P(a_p>a_s).
\label{eq:d_4_1_demo2}
\end{eqnarray}
Since $\gamma_a |h_{ab}|^2$ and $\gamma_c |h_{cb}|^2$ are exponential distributed random variables with parameters $1/\tilde{\gamma}_{ab}$ and $1/\tilde{\gamma}_{cb}$ respectively ($a=p$ or $s$, $c=p$ or $s$, and $b=p,s$ or $r$), then $\forall x \in \mathbf{R}$:
\begin{eqnarray}
\label{eq:liste1}
P(\gamma_a |h_{ab}|^2 \leq \gamma_c |h_{cb}|^2)&=&\frac{\tilde{\gamma}_{cb}}{\tilde{\gamma}_{ab}+\tilde{\gamma}_{cb}} ,\\
\label{eq:liste3}
P(\gamma_a |h_{ab}|^2 \geq (\gamma_c |h_{cb}|^2+1) x)&=&\frac{{\tilde{\gamma}_{ab}}e^{-\frac{x}{\tilde{\gamma}_{ab}}}}{{\tilde{\gamma}_{ab}}+x {\tilde{\gamma}_{cb}}},
\end{eqnarray}
and the probability density function (pdf) of $X_{ab}=\frac{\gamma_a |h_{as}|^2}{\gamma_b |h_{bs}|^2+1}$ ($a=s$ or $r$; $b=p$ or $r$) is given by:
\begin{equation}
\label{eq:pdf_quotient}
f_{X_{ab}}(x)=\frac{e^{-\frac{x}{\tilde{\gamma}_{as}}}}{\tilde{\gamma}_{as}+x \tilde{\gamma}_{bs}}\left( 1+ \frac{\tilde{\gamma}_{as}\tilde{\gamma}_{bs}}{\tilde{\gamma}_{as}+x \tilde{\gamma}_{bs}} \right),\; x \geq 0.
\end{equation}
Since $\frac{1+\frac{\tilde{\gamma}_{rs}^{(s)}\tilde{\gamma}_{ps}}{\tilde{\gamma}_{rs}^{(s)}+y \tilde{\gamma}_{ps}}}{\tilde{\gamma}_{ss}+y \tilde{\gamma}_{rs}^{(p)}} \leq \frac{1+\tilde{\gamma}_{ps}}{\tilde{\gamma}_{ss}}$, $\forall y\geq 0$ and using (\ref{eq:pdf_quotient}), we get:
\begin{equation}
\label{eq:upPEp}
P(a_p>a_s)\leq -\phi_i\left( -\frac{1}{\tilde{\gamma}_{ps}}- \frac{\tilde{\gamma}_{rs}^{(s)}}{\tilde{\gamma}_{ss}\tilde{\gamma}_{ps}} \right)\left( \frac{1+\tilde{\gamma}_{ps}}{\tilde{\gamma}_{ss}} \right)e^{ \frac{1}{\tilde{\gamma}_{ps}} + \frac{\tilde{\gamma}_{rs}^{(s)}}{\tilde{\gamma}_{ss}\tilde{\gamma}_{ps}} }.
\end{equation}

Using (\ref{eq:liste1}),(\ref{eq:liste3}) and (\ref{eq:upPEp}) in (\ref{eq:d_4_1_demo}), we obtain (\ref{eq:occurrence_D4_1}).

\section{Proof of Lemma.\ref{Lemma1}}
From (\ref{eq:outage_prob_CR_help_PU_D1_0}), $\omega$, $\omega_1$
and $\omega_2$ are exponential random variables with parameters $1/\tilde{\gamma}_{rp}$,
$1/\tilde{\gamma}_{sp}$ and
$1/\tilde{\gamma}_{pp}$ respectively. Thus, the
pdfs of $\omega$ and
$\phi=\omega_2-\Lambda_p \omega_1$ are:
\[
\nonumber f_\omega(\omega)=\frac{e^{-\frac{\omega}{\tilde{\gamma}_{rp}}}}{\tilde{\gamma}_{rp}},\omega \geq 0;\;\;\nonumber {{f_\phi}(\phi) = \left\{ {\begin{array}{*{20}{c}}
  {\frac{{{e^{ - \frac{\phi}{{{\tilde{ \gamma }_{pp}}}}}}}}{{{\tilde{ \gamma }_{pp}} + {\Lambda _p} \tilde{\gamma}_{{sp}}}},}&{\phi \geq 0} \\
  {\frac{{{e^{\frac{\phi}{{{\Lambda _p}\tilde \gamma_{{sp}}}}}}}}{{{\tilde{ \gamma }_{pp}} + {\Lambda _p} \tilde{\gamma}_{{sp}}}},}&{\phi \leq 0.}
\end{array}} \right.}
\]
Finally, we obtain the conditional primary outage probability:
\begin{eqnarray}
\label{eq:p_out_demo_CR_help_PU} \nonumber  P(z<\Lambda_p)& = & \int_{ -
\infty
}^{{\Lambda _p}} {{f_z}(z)dz}=\int_{ - \infty }^{{0}} {\frac{{{\Lambda _p}\tilde{\gamma}
{_{sp}}{e^{\frac{z}{{{\Lambda _p}\tilde{\gamma}
{_{sp}}}}}}}}{{{\beta _1}{\beta _2}}}}dz\\
\nonumber &+& \int_{0 }^{{\Lambda_p}} {\frac{{{\Lambda _p}\tilde{\gamma}
{_{sp}}}e^{-\frac{z}{\tilde{\gamma}_{rp}^{(p)}}}}{{{\beta _1}{\beta _2}}} + \frac{{{\tilde{\gamma} _{pp}}\left( {{e^{ - \frac{z}{{{\tilde{\gamma} _{rp}^{(p)}}}}}} -
{e^{ - \frac{z}{{{\tilde{\gamma} _{pp}}}}}}}
\right)}}{{{\beta _1}\left( {{\tilde{\gamma} _{rp}^{(p)}} - {\tilde{\gamma}
_{pp}}} \right)}}  } dz\\
& = & \lambda_1+\lambda_2,
\end{eqnarray}
where $z=\omega+\phi$, $f_z(z)$ is its pdf, $\beta_1=\tilde{\gamma}_{pp}+\Lambda_p
\tilde{\gamma}_{sp}$, $\beta_2=\tilde{\gamma}_{rp}^{(p)}+\Lambda_p \tilde{\gamma}_{sp}$; $\lambda_1$ and $\lambda_2$ are given by (\ref{eq:d1}) and (\ref{eq:d2}) respectively. This completes the proof of Lemma.\ref{Lemma1}.

\section{Proof of Lemma.\ref{Lemma2}}
Using (\ref{eq:pdf_quotient}) for $X_{sb}$ ($b=p$ or $r$), we obtain:
\begin{eqnarray}
\label{eq:prob_sec}
&P&\left( X_{sp} + X_{sr} < \Lambda_s \right)=\\
\nonumber &=& \int_{0}^{\Lambda_s}{\frac{\left( 1+ \frac{\tilde{\gamma}_{ss}\tilde{\gamma}_{rs}^{(p)}}{\tilde{\gamma}_{ss}+x \tilde{\gamma}_{rs}^{(p)}}  \right)}{\tilde{\gamma}_{ss}+x \tilde{\gamma}_{rs}^{(p)}}} \left( e^{-\frac{x}{\tilde{\gamma}_{ss}}}- \frac{\tilde{\gamma}_{ss}e^{-\frac{\Lambda_s}{\tilde{\gamma}_{ss}}}}{\tilde{\gamma}_{ss}+ (\Lambda_s-x) \tilde{\gamma}_{ps}} \right)dx.
\end{eqnarray}
Since ${\frac{\left( 1+ \frac{\tilde{\gamma}_{ss}\tilde{\gamma}_{rs}^{(p)}}{\tilde{\gamma}_{ss}+x \tilde{\gamma}_{rs}^{(p)}}  \right)}{\tilde{\gamma}_{ss}+x \tilde{\gamma}_{rs}^{(p)}}}\leq \frac{\tilde{\gamma}_{rs}^{(p)}+1}{\tilde{\gamma}_{ss}}$, then using this upper bound in (\ref{eq:prob_sec}), we obtain (\ref{eq:conditional_sec_D1}).
This completes the proof of Lemma.\ref{Lemma2}.

\bibliographystyle{IEEEtran}
\bibliography{IEEEabrv,tau}

\begin{thebibliography}{1}
\providecommand{\url}[1]{#1}
\csname url@samestyle\endcsname
\providecommand{\newblock}{\relax}
\providecommand{\bibinfo}[2]{#2}
\providecommand{\BIBentrySTDinterwordspacing}{\spaceskip=0pt\relax}
\providecommand{\BIBentryALTinterwordstretchfactor}{4}
\providecommand{\BIBentryALTinterwordspacing}{\spaceskip=\fontdimen2\font plus
\BIBentryALTinterwordstretchfactor\fontdimen3\font minus
  \fontdimen4\font\relax}
\providecommand{\BIBforeignlanguage}[2]{{%
\expandafter\ifx\csname l@#1\endcsname\relax
\typeout{** WARNING: IEEEtran.bst: No hyphenation pattern has been}%
\typeout{** loaded for the language `#1'. Using the pattern for}%
\typeout{** the default language instead.}%
\else
\language=\csname l@#1\endcsname
\fi
#2}}
\providecommand{\BIBdecl}{\relax}
\BIBdecl

\bibitem{Mitola}
J.~Mitola and G.~Q.~J. Maguire, ``Cognitive radio: making software radios more
  personal,'' \emph{IEEE Personal Commun.}, vol.~6, no.~4, pp. 13 --18, aug.
  1999.

\bibitem{Haykin}
S.~Haykin, ``Cognitive radio: brain-empowered wireless communications,''
  \emph{IEEE J. Select. Areas Commun.}, vol.~23, no.~2, pp. 201 --220, feb.
  2005.

\bibitem{Laneman1}
J.~Laneman, D.~Tse, and G.~Wornell, ``Cooperative diversity in wireless
  networks: Efficient protocols and outage behavior,'' \emph{{IEEE} Trans. Inf.
  Theory}, vol.~50, no.~12, pp. 3062 --3080, Dec. 2004.

\bibitem{Jaafar_WCM}
W.~Jaafar, W.~Ajib, and D.~Haccoun, ``On the performance of multi-hop wireless
  relay networks,'' \emph{Wireless Communications and Mobile Computing}, Dec.
  2011, [Online] Available: http://www.wileyonlinelibrary.com.

\bibitem{Yang_2009}
Y.~Han, A.~Pandharipande, and S.~Ting, ``Cooperative decode-and-forward
  relaying for secondary spectrum access,'' \emph{{IEEE} Trans. Wireless
  Commun.}, vol.~8, no.~10, pp. 4945 --4950, october 2009.

\bibitem{Zou}
Y.~Zou, J.~Zhu, B.~Zheng, and Y.-D. Yao, ``An adaptive cooperation diversity
  scheme with best-relay selection in cognitive radio networks,'' \emph{{IEEE}
  Trans. Signal Process.}, vol.~58, no.~10, pp. 5438 --5445, oct. 2010.

\bibitem{Yener}
K.~Lee and A.~Yener, ``Outage performance of cognitive wireless relay
  networks,'' in \emph{IEEE Global Telecommun. Conf.,}, dec. 2006, pp. 1--5.

\bibitem{Jaafar_Globecom}
\BIBentryALTinterwordspacing
W.~Jaafar, W.~Ajib, and D.~Haccoun, ``A novel relay-aided transmission scheme
  in cognitive radio networks,'' in \emph{IEEE Global Telecommun. Conf.}, dec.
  2011. [Online]. Available: \url{http://arxiv.org/abs/1109.2843v1}
\BIBentrySTDinterwordspacing

\bibitem{Tse}
D.~Tse and P.~Viswanath, \emph{Fundamentals of Wireless Communication}.\hskip
  1em plus 0.5em minus 0.4em\relax Cambridge University Press,, 2005.

\end{thebibliography}

\end{document}